\documentclass[aps,prl,twocolumn,groupedaddress,reprint,showpacs]{revtex4-2}
\usepackage{graphicx}
\usepackage{setspace}
\usepackage{amsmath}
\usepackage{amssymb}
\usepackage{color}
\usepackage{array}
\usepackage{subfigure}
\usepackage{hyperref}
\usepackage{float}
\usepackage{lipsum}

\usepackage[all]{xy}
\newcommand{\RN}[1]{%
  \textup{\uppercase\expandafter{\romannumeral#1}}%
}

\begin{document}
\title{Suppression of high-frequency components in off-resonant modulated driving protocols for Rydberg blockade gate}
\author{Yuan Sun}
\email[email: ]{yuansun@siom.ac.cn}
\affiliation{CAS Key Laboratory of Quantum Optics and Center of Cold Atom Physics, Shanghai Institute of Optics and Fine Mechanics, Chinese Academy of Sciences, Shanghai 201800, China}

\begin{abstract}
In the rapid development of cold atom qubit platform, the two-qubit Controlled-PHASE Rydberg blockade gate via off-resonant modulated driving has been making significant progress recently. In pursuit of higher fidelity, faster operation and better robustness, a major upgrade about suppression of high-frequency components in the modulation is called for, and a systematic method has been established here for this purpose. The quintessence of this newly constructed method can be interpreted as filtering out the relatively high frequency ingredients embedded in basis functions to generate the modulation waveforms and then analyzing whether they fulfill the requirement of gate condition. It turns out that appropriate waveforms of two-qubit entangling gate protocols can be successfully established via these frequency-adjusted basis functions, with the help of numerical optimization procedures. Moreover, this timely upgrade version can be further enhanced with adaptions to specific finite Rydberg blockade strength values and dual-pulse technique to overcome residual thermal motion of qubit atoms. Besides theoretical derivations, we also thoroughly investigate the representative modulation patterns, demonstrating the versatility of off-resonant modulated driving method in the design of two-qubit entangling Rydberg blockade gate. 
\end{abstract}
\pacs{32.80.Qk, 03.67.Lx, 42.50.-p, 33.80.Rv}
\maketitle


The cold atom qubit platform \cite{RevModPhys.82.2313, RevModPhys.87.1379} is currently attracting more and more attention as intense efforts have been devoted to its development.  While the scale of atomic qubit arrays \cite{PhysRevA.92.022336, Weiss2581Science, Saffman2019prl} and the single-qubit gate fidelity \cite{PhysRevLett.114.100503, PhysRevLett.121.240501, nature604.457, PhysRevX.12.021027} are advancing constantly, enhancing the fidelity of two-qubit entangling Rydberg blockade gate \cite{nphys1183, PhysRevLett.104.010503} becomes one of the most pressing tasks in this field \cite{PhysRevLett.104.010502, PhysRevApplied.15.054020}. Growing interest has been accumulating in this direction, and recently the importance of off-resonant modulated driving (ORMD) gate protocol manifests itself, where the modulation in the atom-light interaction process is deemed as essential to construct this distinguished style of Controlled-PHASE gate (CPHASE gate) \cite{PhysRevApplied.13.024059}. Such protocols can be designed for both one-photon and two-photon ground-Rydberg transitions or even more complicated driving mechanisms \cite{PhysRevApplied.13.024059, OptEx480513}, with possible extensions to ensemble qubits \cite{PhysRevLett.115.093601}. Although the two-qubit Controlled-Z gate (CZ gate) fidelity $\gtrsim 0.999$ in experimental demonstrations still seems as a challenge for now \cite{nphys1178, PhysRevA.105.042430, Lukin2023arXiv}, advances in highly coherent ground-Rydberg transition \cite{Endres2023arXiv}, atomic qubit array single-site addressing capability \cite{Saffman2019prl} and quantum algorithms tailored for cold atom qubits \cite{nphys41567} spark hope and momentum for further investigations.

One core feature of ORMD protocols is that the atomic wave function begins and finishes at the same quantum state, and then it acquires a non-trivial geometric phase from the time evolution. The conditional phase shift of CPHASE gate derives from whether the Rydberg blockade effect takes place. The modulation of driving processes generally consist of amplitude modulation and frequency modulation \cite{PhysRevApplied.13.024059}. Amplitude modulation leads to time-dependent Rabi frequency strengths, where a smooth pulse starts and ends at zero intensity is usually preferable for experimental implementations. Frequency modulation can be effectively introduced in the form of phase modulation on driving lasers or tuning the energy of Rydberg state by external electric, magnetic or even microwave fields. Aiming at upgrading ORMD protocols to the next level, the main theme of this work is about suppressing high-frequency components in ORMD waveforms. This change has several advantages, in consideration of both experimental and theoretical factors. Particularly, it can avoid the distortion on high-frequency spectrum that occurs in the electro-optical modulation instruments, and help to alleviate some delicate non-adiabatic effects in the time evolution of atomic wave functions caused by the higher-order frequency components of driving fields, which is sometimes even not necessarily straightforward to faithfully reveal in numerical simulations. Meanwhile, it naturally possesses an indispensable role in the foreseeable future of fast CZ gates with gate time much less than 100 ns for the cold atom qubit platform.

The rest of contents are organized as follows. We first establish the routine of suppressing high-frequency components of modulation waveforms in the abstract sense, with respect to the basic principles of ORMD protocols. Next we demonstrate how to practically deploy this innovated upgrade with one-photon and two-photon ground-Rydberg transitions. Furthermore, we work on adaption to certain imperfections that often adversely influence experiments, including finite Rydberg blockade strength and Doppler-induced dephasing caused by cold atoms' residual thermal motion. Throughout the text, the gate fidelity is evaluated by the method of Refs. \cite{JAMIOLKOWSKI1972275, CHOI1975285, PhysRevA.71.062310, PEDERSEN200747}.

It has been well established by the ORMD protocols that the induced changes in the wave functions reasonably satisfy the requirement of two-qubit phase gate if the off-resonant atom-light interaction is appropriately modulated \cite{PhysRevApplied.13.024059, OptEx480513}. The realistic task then becomes to find a rigorous method to systematically compute and characterize the waveforms of modulation. For ease of formal representation and compatibility with numerical calculation, a waveform $w$ can be expressed in terms of linear superpositions of a complete basis $\{g_\nu\}$ for $L^2$ functions on the given finite time interval of gate operations: $w=\sum_{\nu=0}^{\infty} \alpha_\nu g_\nu$ with coefficients $\alpha_\nu$'s. Theoretically it provides the tool of retrieving all possible solutions, while practically the waveforms with truncations in the expansion are sought for that can lead to high-fidelity performance. The choice of basis is certainly not unique, such as the Bernstein polynomials or Fourier series. The major upgrade in pursuit here focuses on waveforms with high-frequency components suppressed, where we assume that the requested high-frequency-suppressing rule is represented by an operator $S_f$. Suppose that $S_f$ has the linearity property, that is, for two waveforms $w_1, w_2$, $S_f(\alpha_1 w_1 + \alpha_2 w_2)= \alpha_1S_f(w_1) + \alpha_2S_f(w_2), \forall \alpha_1, \alpha_2 \in \mathcal{R}$. In particular, it includes the operation of eliminating all higher order frequency components above certain prescribed cut-off value. Then it is straightforward to observe that $S_f(w)=\sum_{\nu=0}^{\infty} S_f(g_\nu)$, which indicates in principle the frequency-suppressing operation of interest is equivalent to construct waveforms via basis functions with necessary adjustments on the frequency domain. In other words, suppression of high-frequency components embedded in the waveform can be translated as the corresponding suppression operations on the basis functions.

\begin{figure}[h]
\centering
\includegraphics[width=0.432\textwidth]{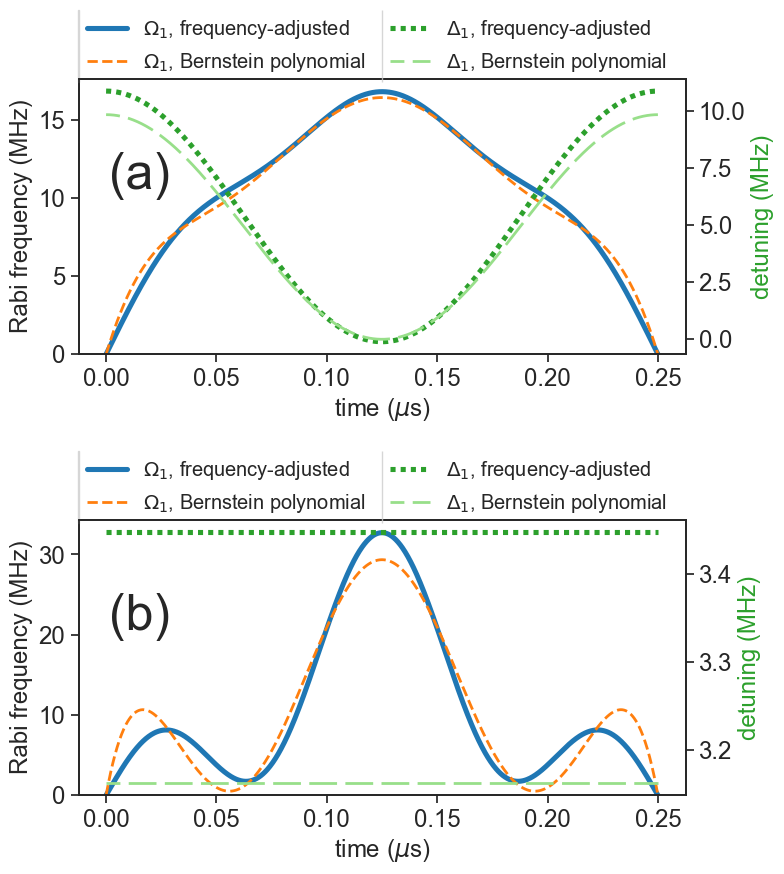}
\caption{(Color online) CZ gate waveforms for one-photon transition. (a) Amplitude and frequency are both modulated. (b) Only amplitude modulation. The calculated gate errors of frequency-adjusted waveforms are less than $10^{-4}$.}
\label{fig1:waveforms_1photon}
\end{figure}

\begin{align}
&\begin{bmatrix} v_1\\v_2\\v_3\\v_4\\v_5\\v_6\end{bmatrix}
=
\begin{bmatrix}
0.1273 & 0.2079 & 0.1310 & 0.0682 \\
0.1806 & 0.1793 & 0.0282 &-0.0085 \\
0.2242 & 0.0892 &-0.0416 &-0.0191 \\
0.2566 &-0.0210 &-0.0386 &-0.0025 \\
0.2766 &-0.1085 & 0.0039 & 0.0028 \\
0.2833 &-0.1416 & 0.0272 & 0.0009 
\end{bmatrix} \cdot
\begin{bmatrix} \sin \pi x \\ \sin 3\pi x \\ \sin 5\pi x \\ \sin 7\pi x\end{bmatrix},\nonumber\\
&\begin{bmatrix} u_1\\u_2\\u_3\\u_4\\u_5\end{bmatrix}
=
\begin{bmatrix}
0.1710 & 0.2316 & 0.1158 \\
0.2382 & 0.1568 & -0.0013\\
0.2887 & 0.0220 & -0.0408\\
0.3199 & -0.0986& -0.0081\\
0.3305 & -0.1463& 0.0173
\end{bmatrix} \cdot
\begin{bmatrix} \sin \pi x \\ \sin 3\pi x \\ \sin 5\pi x \end{bmatrix}.
\label{eq:waveform_v31}
\end{align}

As a concrete step towards applications, we choose to work on frequency-adjusted version of Bernstein polynomials as the basis for amplitude modulation waveforms. To retain the helpful property of a Bernstein polynomial $b$ that starts and ends at zero, it is firstly extended as an odd function on $[-1, 1]$ and subsequently expanded into Fourier series. The specific choice of $S_f$ here is to abandon all higher order terms above a threshold. To keep discussions succinct, only two sets of frequency-adjusted functions $\{u_j\}, \{v_j\}$ defined on $[0, 1]$ will be employed for symmetric waveforms as in Eq. \eqref{eq:waveform_v31} with four significant digits after the decimal point. $u_j$ corresponds to $(b_{j, 10}+b_{10-j, 10})$ and $v_j$ corresponds to $(b_{j, 12}+b_{10-j, 12})$, where $b_{\nu, n}$ represents the $\nu$th Bernstein polynomial of degree $n$. 

Denote the qubit register states as $|0\rangle, |1\rangle$, the Rydberg state as $|r\rangle$ and the singly-excited Rydberg state as $|\tilde{r}\rangle = (|r1\rangle+|1r\rangle)/\sqrt{2}$ assuming the qubit atoms receives the same laser driving. There exist the single-excitation process of $|01\rangle\leftrightarrow|0r\rangle$, $|10\rangle\leftrightarrow|r0\rangle$ and the double-excitation process of $|11\rangle \leftrightarrow |\tilde{r}\rangle$. Without loss of generality, for one-photon transitions the single-excitation process can be described as:
\begin{eqnarray}
\label{eq:Hamiltonian_1photon_1body}
H_{s1}/\hbar = &\frac{1}{2}\Omega_1|r0\rangle\langle 10| + \frac{1}{2}\Omega_1|0r\rangle\langle 01| + \text{H.c.} \nonumber\\
&+ \Delta_1 |r0\rangle\langle r0| + \Delta_1 |0r\rangle\langle 0r|,
\end{eqnarray}
with Rabi frequency $\Omega_1(t)$ and detuning $\Delta_1(t)$. Meanwhile, the double-excitation process with the idealized Rydberg blockade effect can be described as: 
\begin{equation}
\label{eq:Hamiltonian_1photon_2body}
H_{s2}/\hbar = \frac{1}{\sqrt{2}}\Omega_1|\tilde{r} \rangle \langle 11| + \text{H.c.}
+ \Delta_1 |\tilde{r}\rangle\langle \tilde{r}|.
\end{equation}

According to the abstract principles of designing OMRD protocols, CZ gate waveforms can be numerically computed with respect to interactions governed by $H_{s1}, H_{s2}$. In particular, waveforms are firstly constructed via Bernstein polynomials, which then serve as the starting point of optimization procedures to acquire frequency-adjusted versions. Fig. \ref{fig1:waveforms_1photon} shows typical results with idealized Rydberg blockade, where the frequency-adjusted waveforms are $\Omega_1(x)/2\pi = 9.71u_1(x)+13.55u_2(x)+0.10u_3(x)+26.29u_4(x)+8.89u_5(x) \text{ MHz}, \Delta_1(x)/2\pi=5.358+5.497\cos 2\pi x \text{ MHz}$ for (a) and $\Omega_1(x)/2\pi = 42.20u_1(x)-24.93u_2(x)-25.00u_3(x)-42.00u_4(x)+111.85u_5(x) \text{ MHz}, \Delta_1/2\pi =3.448 \text{ MHz}$ for (b) with $x=t/T_p$ and pulse time $T_p$ as 250 ns.

\begin{figure}[h]
\centering
\includegraphics[width=0.432\textwidth]{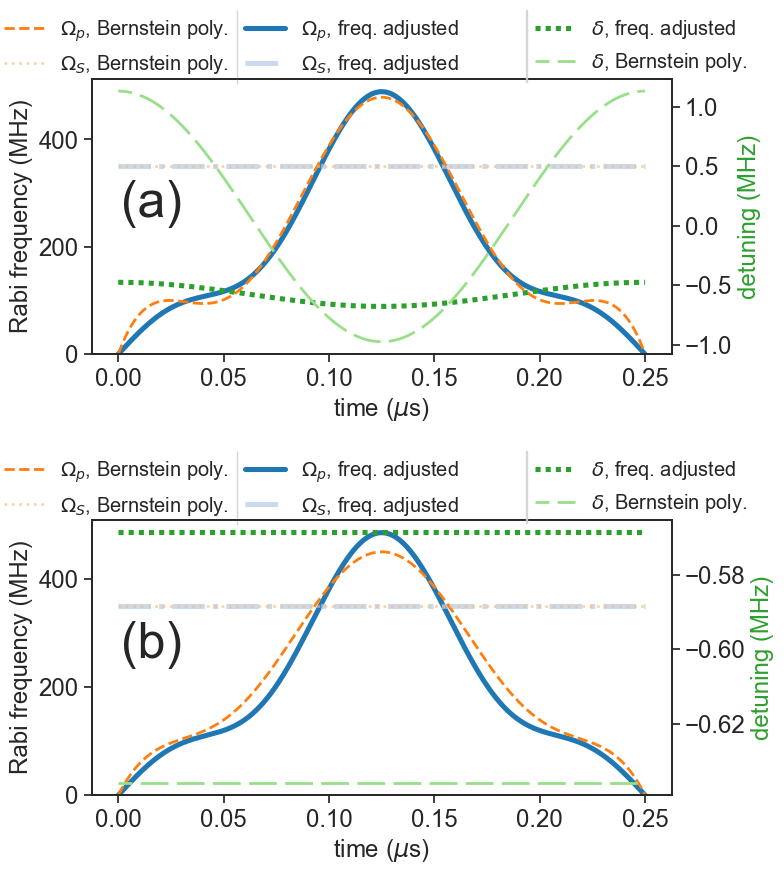}
\caption{(Color online) CZ gate waveforms for two-photon transition. (a) Amplitude and frequency are both modulated. (b) Only amplitude modulation. The calculated gate errors of frequency-adjusted version are smaller than $10^{-4}$.}
\label{fig2:waveforms_2photon}
\end{figure}

On the other hand, for typical two-photon transitions \cite{PhysRevApplied.15.054020, PhysRevA.105.042430, OptEx480513}, the single-excitation process can be formulated as $H_{d1}/\hbar = \frac{\Omega_p}{2}|10\rangle\langle e0| + \frac{\Omega_S}{2}|e0\rangle\langle r0| + \frac{\Omega_p}{2}|01\rangle\langle 0e| + \frac{\Omega_S}{2}|0e\rangle\langle 0r| + \text{H.c.} + \Delta|e0\rangle\langle e0| + \delta|r0\rangle\langle r0| + \Delta|0e\rangle\langle 0e| + \delta|0r\rangle\langle 0r|$ with one-photon detuning $\Delta$ and two-photon detuning $\delta$, while the double-excitation process of $|11\rangle$ with idealized Rydberg blockade can be formulated as $H_{d2}=\frac{\Omega_p}{\sqrt{2}}|11 \rangle \langle \tilde{e}| + \frac{\Omega_S}{2} |\tilde{e}\rangle\langle \tilde{r}| + \frac{\Omega_p}{2} |\tilde{r}\rangle\langle \tilde{R}| + \text{H.c.} + \Delta |\tilde{e}\rangle\langle \tilde{e}| + \delta |\tilde{r}\rangle\langle \tilde{r}|+ (\Delta+\delta) |\tilde{R}\rangle\langle \tilde{R}|$, with $|\tilde{e}\rangle = (|e1\rangle+|1e\rangle)/\sqrt{2}, |\tilde{R}\rangle = (|re\rangle+|er\rangle)/\sqrt{2}$.

Not surprisingly, it turns out CZ gate waveforms with suppression of high-frequency components can be successfully obtained for two-photon transitions pertinent to $H_{d1}, H_{d2}$, and Fig. \ref{fig2:waveforms_2photon} shows typical results, where the frequency-adjusted waveforms are $\Omega_p/2\pi = 270.84u_1-25.68u_2-88.66u_3-21.08u_4+1021.0u_5 \text{ MHz}, \delta/2\pi=-0.577+0.101\cos 2\pi x \text{ MHz}$ for (a) and $\Omega_p/2\pi = 260.85u_1-11.51u_2-79.85u_3+0.0u_4+992.41u_5 \text{ MHz}, \delta/2\pi =-0.636 \text{ MHz}$ for (b), both with $\Omega_S/2\pi=350 \text{ MHz}, \Delta/2\pi=5000 \text{ MHz}$. There are many other styles of amplitude modulation \cite{OptEx480513} and the frequency-adjusted version is applicable to all of them. 

\begin{figure}[h]
\centering
\includegraphics[width=0.432\textwidth]{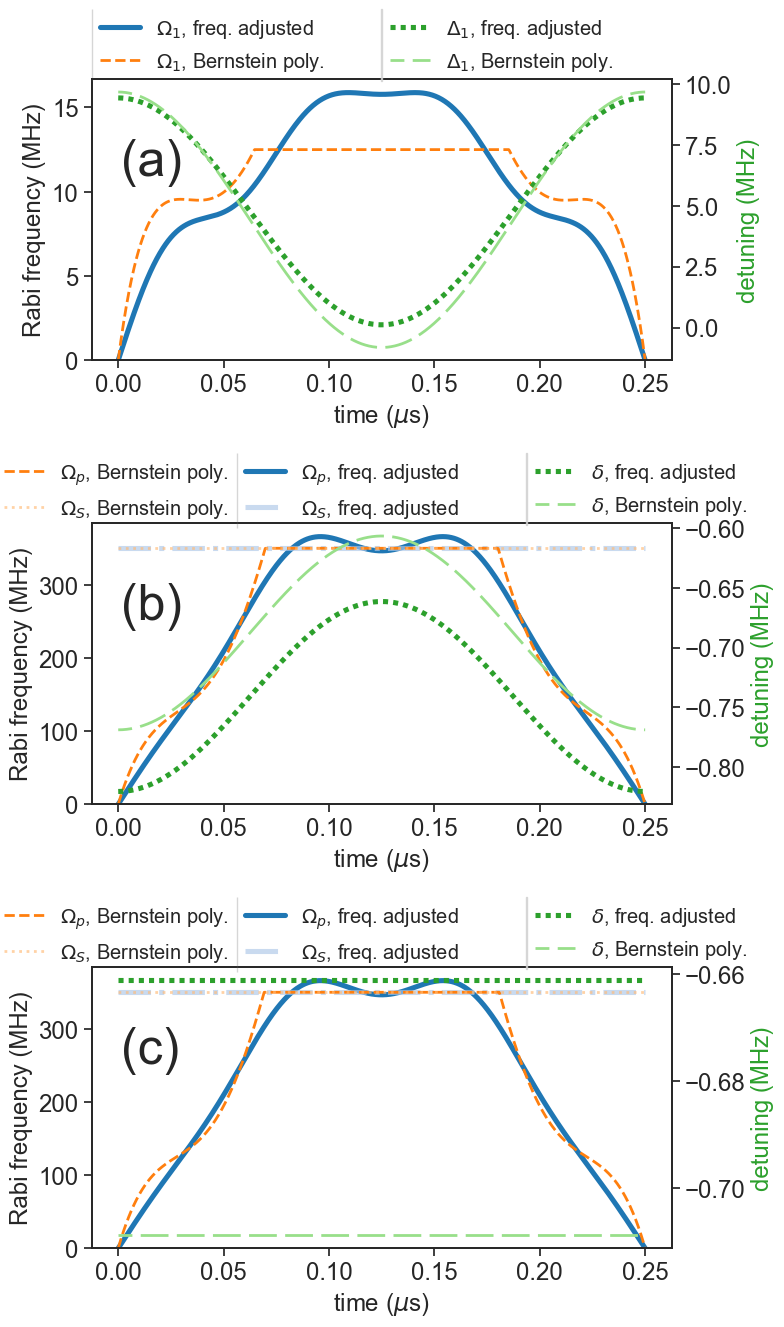}
\caption{(Color online) CZ gate waveforms with imposed limit on maximum values of Rabi frequency amplitudes. (a) One-photon transition. (b) Two photon transition with both amplitude and frequency modulations. (c) Two photon transition with only amplitude modulation. The calculated gate errors of frequency-adjusted version are smaller than $10^{-4}$.}
\label{fig3:limit_RabiFreq}
\end{figure}

Sometimes the peak power of driving lasers faces harsh constraints in experiments and then reducing maximum values of Rabi frequency amplitudes becomes an intriguing task of gate protocols. The design of ORMD waveforms is capable of fulfilling this requirement without loss of fidelity or comprises of robustness. Specifically speaking, ORMD waveforms can properly function with an imposed upper limit of Rabi frequency amplitude, with or without frequency modulations. Moreover, we discover that it is possible to apply the systematic method as discussed in previous paragraphs to suppress high-frequency components in such waveforms. For instance, waveforms expressed by Bernstein polynomials $b_{\nu, n}$ with imposed upper limit can be approximated by linear superposition of $b_{\nu, m}$ higher degree $m>n$, which can then be devolved to high-frequency-suppressing process. Nevertheless, as the laser technology advances rapidly nowadays, it is worthwhile to note that reduction of laser power is part of the more integrated strategy for better gate performance of ORMD protocols rather than a standalone requirement.

Fig. \ref{fig3:limit_RabiFreq} presents typical waveforms demonstrating these concepts, where the frequency-adjusted waveforms are $\Omega_1/2\pi = 16.41v_1 -0.49v_2 +4.06v_3 +22.38v_4 +29.25v_5 -2.37v_6 \text{ MHz}, \Delta_1/2\pi=4.772 +4.660\cos 2\pi x \text{ MHz}$ for (a),  $\Omega_p/2\pi = 203.25v_1 -5.91v_2 -0.96v_3 +1199.83v_4 +156.87v_5 +0.0v_6 \text{ MHz}, \delta/2\pi=-0.741 -0.079\cos 2\pi x \text{ MHz}$ for (b) and $\Omega_p/2\pi = 200.45v_1 -4.29v_2 +3.14v_3 +1196.44v_4 +157.56v_5 +0.0v_6 \text{ MHz}, \delta/2\pi = -0.709 \text{ MHz}$ for (c) with $\Omega_S/2\pi=350 \text{ MHz}, \Delta/2\pi=5000 \text{ MHz}$.

\begin{figure}[h]
\centering
\includegraphics[width=0.432\textwidth]{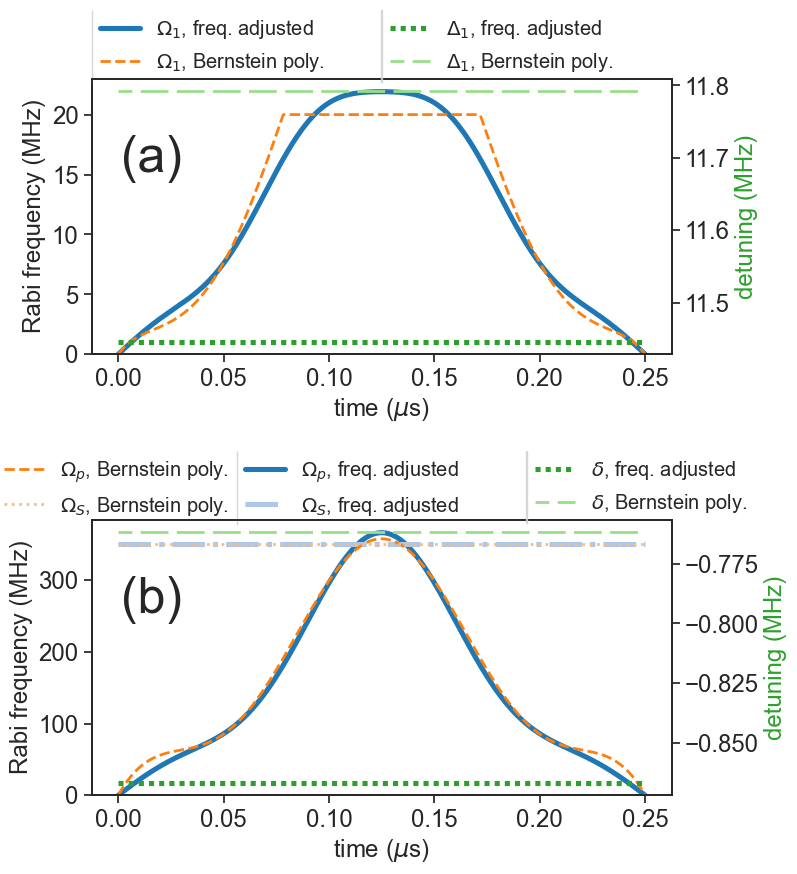}
\caption{(Color online) CZ gate waveforms of dual-pulse technique. (a) One-photon transition. (b) Two photon transition. The calculated gate errors of frequency-adjusted version are smaller than $10^{-4}$.}
\label{fig4:dual_pulse}
\end{figure}

The residual thermal motion of the qubit atoms can cause Doppler-induced dephasing \cite{Saffman2019prl, PhysRevApplied.15.054020} in two-qubit gate interactions. Within the framework of ORMD protocols, usually this type of deviations from the ideal situation leads to first-order effect in phase and second-order effect in population, and if two identical pulses are applied consecutively but in completely opposite directions, their first-order effect in the phase cancels out, which brings about the dual-pulse technique \cite{PhysRevApplied.13.024059}. Fortunately, the method of high-frequency-suppressing turns out to be compatible with the dual-pulse technique as well, including multiple modulation styles. Fig. \ref{fig4:dual_pulse} shows typical results, where the frequency-adjusted waveforms are $\Omega_1/2\pi = 8.39v_1 -0.12v_2 -10.0v_3 +49.24v_4 +29.25v_5 +0.0v_6 \text{ MHz}, \Delta_1/2\pi=11.446 \text{ MHz}$ for (a) and $\Omega_p/2\pi = 118.41u_1 +52.34u_2 -99.12u_3 +195.13u_4 +602.87u_5 \text{ MHz}, \delta/2\pi =-0.867 \text{ MHz}$ with $\Omega_S/2\pi=350 \text{ MHz}, \Delta/2\pi=5000 \text{ MHz}$ for (b).

\begin{figure}[h]
\centering
\includegraphics[width=0.432\textwidth]{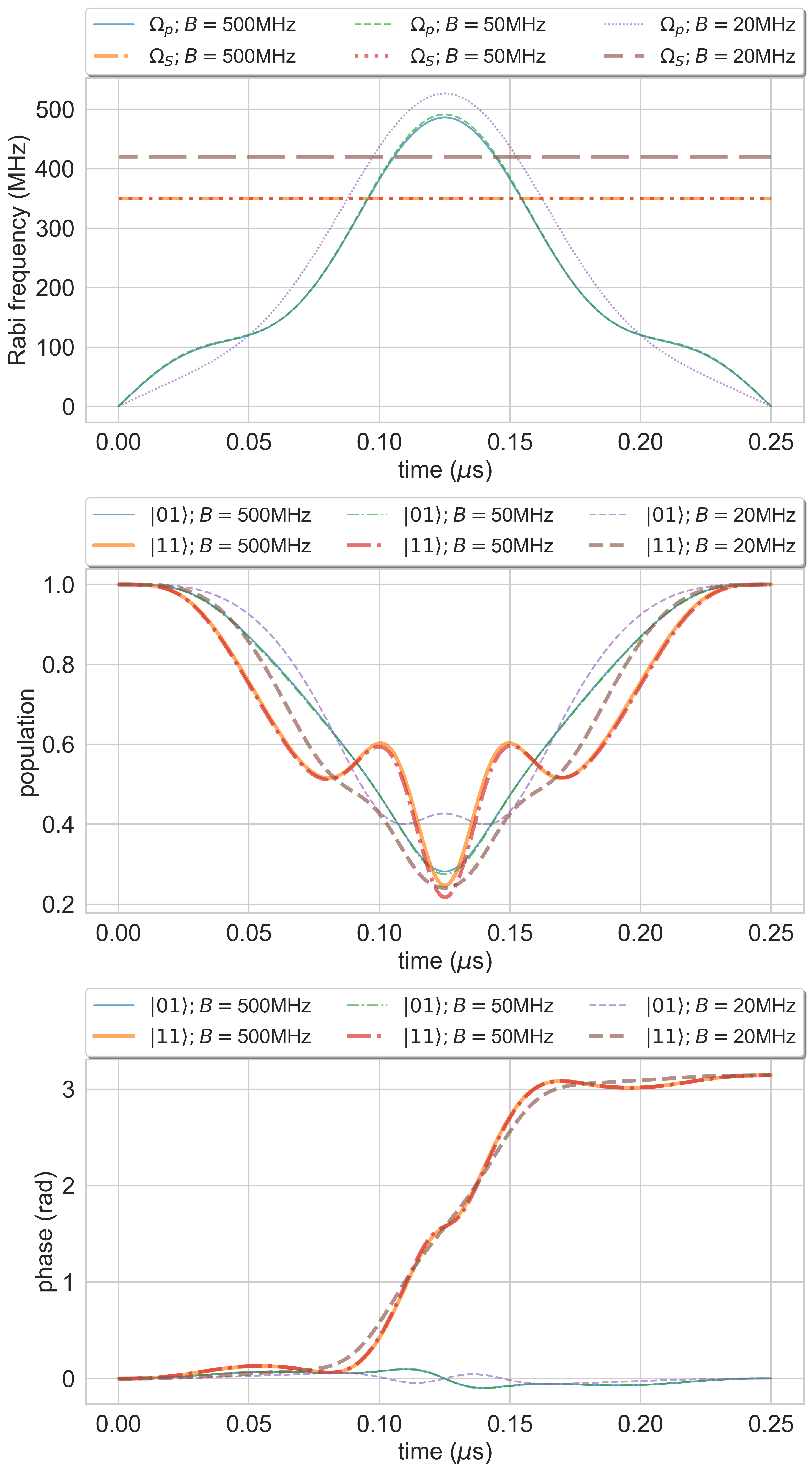}
\caption{(Color online) Frequency-adjusted CZ waveforms of ORMD with RISA, whose calculated gate errors $<10^{-4}$. $\delta_q=0$ for all three cases here.}
\label{fig5:ORMD_withRISA}
\end{figure}

The fundamental structure of Rydberg dipole-dipole interaction is far more complicated beyond the idealized Rydberg blockade effect \cite{RevModPhys.82.2313}, and ORMD with Rydberg interaction strength adaption (RISA) has been devised to deal with finite Rydberg blockade strength \cite{OptEx480513}. According to the single-channel F\"{o}rster resonance model \cite{PhysRevA.96.042306} of $|rr\rangle \leftrightarrow |qq'\rangle$ with the coupling strength as $B$ and the small energy penalty term as $\delta_q$ for $|qq'\rangle$, the additional Hamiltonian term can be expressed by $H_{sF}=\frac{1}{\sqrt{2}}\Omega_1|rr\rangle \langle \tilde{r}| + B|qq'\rangle \langle rr| + \text{H.c.} + 2\Delta_1 (|rr\rangle\langle rr|+|qq'\rangle\langle qq'|) + \delta_q |qq'\rangle\langle qq'|$ for single-photon transition's $H_{s2}$ and $H_{dF}=\frac{\sqrt{2}\Omega_S}{2} |\tilde{R}\rangle \langle rr| + B|rr\rangle \langle qq'| + \text{H.c.} + 2\delta(|rr\rangle\langle rr|+|qq'\rangle\langle qq'|) + \delta_q|qq'\rangle\langle qq'|$ for two-photon transition's $H_{d2}$. These extra complexities can be harmonically incorporated into the design process even many more states are involved. 

In fact, ORMD with RISA performs particularly well with the frequency-adjusted version. High-frequency-suppressing produces extra benefits, possibly including better quality of dark state mechanism, less non-adiabatic effects \cite{PhysRevLett.103.110501, PhysRevA.90.033408} and smaller population leakage, an illustrative example of which is shown in Fig. \ref{fig5:ORMD_withRISA}. This progress helps to clarify the previous theoretically elusive point of whether very large $B$ is required for Rydberg blockade CZ gate to reach high fidelity. For relatively smaller value of $B$, non-negligible population is transferred to the two-body doubly-excited Rydberg states inevitably, which also receive a dressing on the order of $\sqrt{B^2 + \delta_q^2}/2$, but ORMD with RISA can take most of the moved population back with appropriate waveforms.

\begin{figure}[h]
\centering
\includegraphics[width=0.5\textwidth]{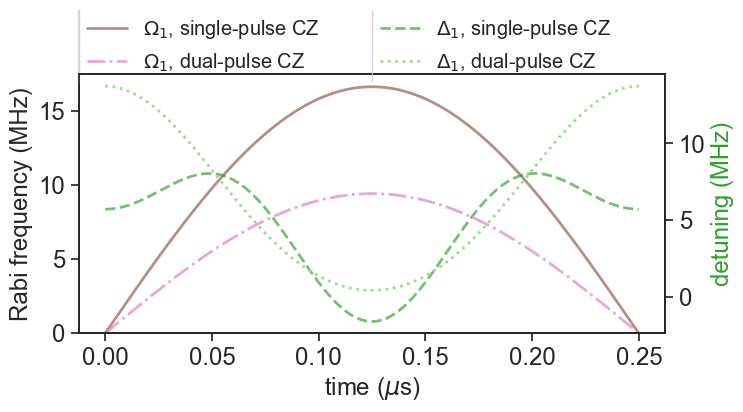}
\caption{(Color online) Waveforms with amplitude modulation of only one frequency element, where $\sin \pi x$ is the lowest order term of Fourier series on $[-1, 1]$. More specifically, the single-pulse waveform is $\Omega_1/2\pi = 16.62 \sin \pi x \text{ MHz}, \Delta_1/2\pi=4.749 + 3.652\cos 2\pi x -2.677\cos 4\pi x \text{ MHz}$ and the dual-pulse waveform is $\Omega_1/2\pi = 9.41 \sin \pi x \text{ MHz}, \Delta_1/2\pi=6.609 + 6.641\cos 2\pi x +0.476\cos 4\pi x \text{ MHz}$. The calculated gate errors are smaller than $10^{-4}$.}
\label{fig6:one_freq_amplitude}
\end{figure}

Following the procedures established so far, pushing towards waveforms consisting of even lower frequency modulation in Rabi frequency amplitude and detuning terms is certainly attainable, with such an example shown in Fig. \ref{fig6:one_freq_amplitude}. One may contrive to get a solution with lowest possible frequency components under special cases, but considering the finite strengths of Rydberg blockade effect, suppressing scattering or leakage and many other theoretical and technical difficulties, such a solution won't necessarily become the optimal choice. From the viewpoint of completeness in designing ORMD protocols, the high-frequency-suppressing method introduced here is equivalent to truncation in Fourier series. Nevertheless, it is noteworthy that unphysical results with Rabi frequency amplitude somewhere smaller than 0 need to be neglected in numerical practices.

In conclusion, the design of ORMD protocols has been significantly enhanced with the fresh upgrade of suppressing high-frequency components embedded in the waveforms, with respect to two-qubit Rydberg blockade gate. According to the obtained results, the ORMD protocols can properly function with satisfactory performance without the involvement of relatively higher frequency ingredients. Following the principles of this essential improvement, a rich variety of ORMD waveforms can be derived for one-photon and two-photon ground-Rydberg transitions, including the special considerations for limiting peak laser intensity, dual-pulse technique and finite Rydberg blockade strength. These findings will become the quintessence for the next stage of experimental efforts pursuing high-fidelity and fast two-qubit gates in the cold atom qubit platform. Furthermore, the high-frequency-suppressing method is applicable to other types of gate protocols. Last but not least, the concept of this work can be extended to construct two-qubit controlled-PHASE gate with arbitrarily designated phase and multi-qubit gate as well, such as the three-qubit Toffoli gate.

\begin{acknowledgements}
This work is supported by the National Natural Science Foundation of China (Grant No. 92165107 and No. 12074391) and the Chinese Academy of Sciences. The author thanks Ning Chen, Xiaodong He, Zhirong Lin, Peng Xu and Hui Yan for many in-depth discussions.
\end{acknowledgements}

\bibliography{sackbut_ref}

\end{document}